\documentclass[aps,pre,twocolumn]{revtex4-1}

\usepackage{amsmath,amssymb}
\usepackage{graphicx}

\begin{document}

%\preprint{}

\title
{%Promoters regulated by
%repressors with fast binding kinetics can 
  %have large transcriptional bursts
  Dynamic competition between transcription initiation and
 repression:  Role of non-equilibrium steps in cell to
  cell heterogeneity
}

\author{Namiko Mitarai}
\email[]{mitarai@nbi.dk}
\affiliation{
Center for Models of Life, 
Niels Bohr Institute, University of Copenhagen,
Blegdamsvej 17, 2100 Copenhagen, Denmark. }
\author{Szabolcs Semsey}
\affiliation{
Center for Models of Life, 
Niels Bohr Institute, University of Copenhagen,
Blegdamsvej 17, 2100 Copenhagen, Denmark. }
\author{Kim Sneppen}
\email[]{sneppen@nbi.dk}
\affiliation{
Center for Models of Life, 
Niels Bohr Institute, University of Copenhagen,
Blegdamsvej 17, 2100 Copenhagen, Denmark. }


\date{\today}

\begin{abstract}
Transcriptional repression may cause transcriptional noise
by a competition between repressor and RNA polymerase binding. Although promoter activity is
often governed by a single limiting step, we argue here that 
the size of the noise strongly depends on whether this step is the initial equilibrium binding or one of the subsequent uni-directional steps.
Overall, we show that non-equilibrium steps of transcription initiation systematically increase the cell to cell heterogeneity in bacterial populations. In particular, this allows also weak promoters 
to give substantial transcriptional noise.
\end{abstract}

%\keywords{}

\maketitle
\section{Introduction}
Protein production in living cells 
is the result of the combined dynamics 
of transcription and translation, through the activity of
first RNA polymerase (RNAP) 
that synthesizes messenger RNA (mRNA), and subsequently
the ribosomes that translate the information on mRNA to proteins.
Because each mRNA typically are translated many times
\cite{bremer1996modulation},
the fluctuations in protein number are sensitive
to fluctuations in the number of produced mRNA \cite{friedman2006linking,yu2006probing}.
Therefore there have been substantial 
interest in determining the noise in this number \cite{golding,taniguchi,so,sanchez}.
This noise is primarily governed
by the stochastic dynamics of RNAP around the promoters,
which are the regions on the DNA that direct initiation of 
the transcription process.

With recent availability of the technology for counting individual mRNAs in {\sl E. coli} cells \cite{golding,taniguchi,so,sanchez}, 
it has become feasible to quantify 
the interplay between noise in gene expression 
and dynamics around the promoter. The degree of cell-to-cell variability in the number of a given mRNA is often quantified by the Fano factor, the ratio between the variance and the mean. 
The Fano factor exceeds one when the transcription is bursty.  Such transcription burstiness can be obtained from a model where a gene switches between an ``on-state" with high promoter activity and an ``off-state" with low activity \cite{golding,mitarai,nakanishi,so,sanchez,Jones}.  In this simple scenario, {%\bf
  Fano factor gives the estimate of the number of transcripts produced per ``on-state''}, and such a scenario can be realized by different molecular mechanisms.

\begin{figure}[h]
\begin{center}
\vspace{-0.0cm}
\includegraphics[width=1.0\hsize]{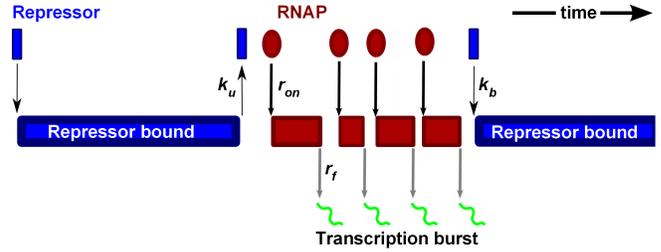}
\caption{Illustration of the competition between
a simple transcriptional repressor (blue) and the RNA polymerase (red)
in terms of the time intervals they occlude the promoter. 
Notice that a bound RNAP takes time to initiate transcription.
It is only when the promoter is open, 
that there is a direct competition for the available space.
The probability that the repressor wins this competition is 
$k_b/(k_b+r_{on})$ where $r_{on}$ is the effective on-rate of the RNAP 
(see Fig. 2) and $k_b$ is the binding rate of the repressor.
The number of times the RNAP binds before the repressor rebinds
is given by $r_{on}/k_b$. 
}\label{figure0}
\end{center}
\end{figure}

Transcriptional regulators influence RNA polymerase (RNAP) access to
promoters, and may cause alternating periods of low and high promoter activity,
depending on the presence or absence of the regulator near the promoter (Fig.~\ref{figure0}).
When the repressor is the source of the burstiness, the measurements of the Fano factor for mRNA levels may allow for quantification of 
the relative sizes of on-rates of transcriptional repressors 
and on-rates of RNAP \cite{nakanishi}.
A recent study \cite{Jones} reported Fano factors in the presence of a transcriptional repressor. The measured dependence of noise on repressor concentration was reproduced by using a one-step model for transcription initiation, assuming that RNAP binding to the promoter sequence is the rate limiting step.

However, transcription initiation in {\sl Escherichia coli} involves 
at least three steps: closed complex, open complex and 
elongation initiation \cite{mcclure,mcclure1980} 
(see Fig.~\ref{modelfig}) 
of which the two later steps are often limiting \cite{ mcclure, spassky, roy, roy2}. Measuring the distributions of time intervals between two subsequent transcription events directly demonstrated that the tetA promoter has at least two limiting steps \cite{Muthukrishnan}.
In cases where promoter activity is limited by later steps of the initiation process, the RNAP is bound to the promoter for a longer period.
This inhibits the access for subsequent RNAPs as well as for 
transcription factors in the occluded region \cite{bendtsen} 
as indicated by the red squares in Fig.~\ref{figure0}. 
{%\bf
  In fact ref. \cite{bendtsen} studied a synthetic model system where
the time RNAP spends on a promoter allowed a four fold
repression of a partly overlapping promoter.
 
Here we analyze how mutual exclusion between transcription factors
and RNAP influences the noise level.}
By taking the multi-step transcription 
initiation explicitly into account, our study emphasizes that although the activity of a promoter may be limited by a single bottleneck process, it does matter whether this 
limiting process is early or late in the 
transcription initiation process.
% Results and Discussion can be combined.
\section{Model}
\begin{figure}[h]
\includegraphics[width=1.0\hsize]{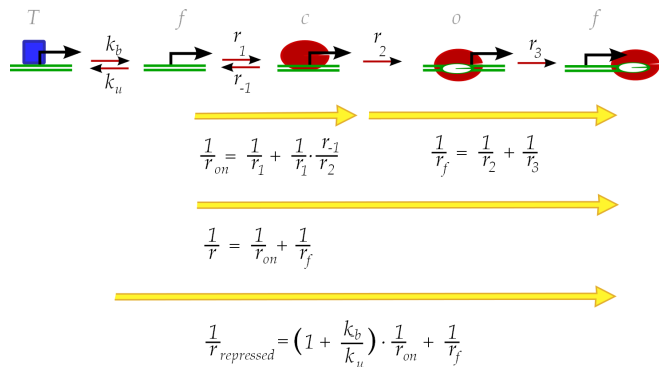}
\caption{
Three-step promoter model of \protect{\cite{mcclure}}
exposed to a repressor.
The appropriate states is marked $T$, $f$,$c$ and $o$,
and the figure illustrate how this 
can be is simplified to a process that focuses
on the difference between the time
$1/r_{on}$ of the RNA polymerase association
and the time consumed by subsequent steps.  
In vitro data for LacUV5 is $r_{1}/r_{-1}=0.16 [RNAP]/nM$  $r_2=0.095/sec$, $r_3=2/sec$ \cite{spassky}
where [RNAP] is free RNA polymerase concentration.}
\label{modelfig}
\end{figure}

Figure~\ref{modelfig} illustrates the interplay between a 
simple transcriptional repressor, acting solely by promoter occlusion, and the activity of the promoter it regulates.
The transcription factor binds to the promoter with a rate $k_b$ when it is free,  and unbind with a rate $k_u$.
We assume the McClure three-step promoter model \cite{mcclure,mcclure1980}  for the transcription initiation. 
The RNAP binds to the free promoter
with a rate $r_1$ to form a closed complex. Subsequently it can
unbind with a rate $r_{-1}$ or form an open complex with a rate
$r_2$. The latter step is a non-equilibrium step, followed by 
a subsequent elongation initiation with a rate $r_3$.

If there is no transcriptional repression,
the total time between subsequent promoter initiations 
can now be obtained by adding together the times 
for the individual steps in the
initiation process. This is illustrated in Fig.~\ref{modelfig}, where this total time $1/r$ is given as the sum
of an effective on-time $1/r_{on}=1/r_1+(1/r_1)\cdot (r_{-1}/r_2)$,
and the time needed for the subsequent step $1/r_f=1/r_2+1/r_3$.
Noticeably this sum rule incorporates all the three standard step of the McClure
promoter model, with the additional caveat that the reversible binding
step takes some additional time because the RNAP may bind and unbind several times before the irreversible open complex is formed. 
The total time between
two transcription initiations is accordingly
\begin{equation}
\frac{1}{r} = \frac{1}{r_{on}} + \frac{1}{r_f}
\end{equation}
where the 0.5-1 seconds time interval 
it take the RNAP to move away from the 
promoter after transcription initiation for simplicity is included 
in $1/r_f$. Therefore, a promoter that is
limited by a small elongation initiation rate $r_f$ 
can have an ``on-rate" $r_{on}$ which is much higher than its overall
initiation rate $r = r_{on} \cdot r_f /(r_{on}+r_f)$ \cite{sneppen2005}.

{%\bf
  It should be noted that such multiple sequential steps
  in promoter initiation can reduce the Fano factor below one,
  if each step takes a similar time scale \cite{mitarai}.
  Obviously the process will be well-approximated 
  by a single Poisson process if one of the steps is the rate limiting,
  which gives the Fano factor one. This is the largest Fano factor
that the system with the sequential reaction steps can achieve.} 

{%\bf
  Repressors make the reaction steps to branch out to a repressed state,
  which allows Fano factor to exceed one. In Fig 2, this branching happens around the ``$f$'' state, with the left branch representing the repressed 
  state (``$T$'' state).}
Noticeably, a repressor that exclusively acts through promoter 
occlusion only interferes with the on-rate $r_{on}$ \cite{nakanishi}.
In other word,  when the RNAP is already on the promoter, then such a 
repressor cannot access the initiation complex 
and influence the subsequent RNAP activity. 
This gives the average initiation time interval under repressor 
$1/r_{repressed}=(1+1/K)\cdot (1/r_{on})+1/r_f$ (Fig.~\ref{modelfig}), 
where the dissociation constant of the repressor $K=k_u/k_b$ quantifies the binding strength of the repressor.
The average mRNA number $\langle m\rangle$ is then given by \cite{nakanishi}
\begin{equation}
\langle m \rangle = \frac{r_{repressed}}{\gamma} = 
\frac{r_{on}/\gamma}{1+R+1/K},
\label{repressm}
\end{equation} 
where the aspect ratio $R=r_{on}/r_f$ characterizes the promoter architecture \cite{sneppen2005}, 
and $\gamma$ is the mRNA degradation rate.

The described reaction scheme  (Fig.~\ref{modelfig}) provides a stochastic mRNA production process. Combined with the
mRNA degradation at a constant rate $\gamma$, the variance of mRNA number in the steady state, 
$\sigma^2=\langle m^2\rangle - \langle m \rangle^2$, can be calculated by using the 
master equations.  We performed the calculation for both the full three-step  initiation model 
and the effective two-step initiation model described by  the irreversible binding with $r_{on}$ and subsequent elongation with a rate $r_f$
\cite{nakanishi}. The detail of the derivation is given in appendix \ref{fanoderive}.
We focus on the Fano factor $\nu = \sigma^2/\langle m\rangle$ as a measure 
of the cell to cell heterogeneity, 
which should be one if the mRNA production is a single step Poisson process.

\section{Results}
\begin{figure}[h]
 \includegraphics[width=1.0\hsize]{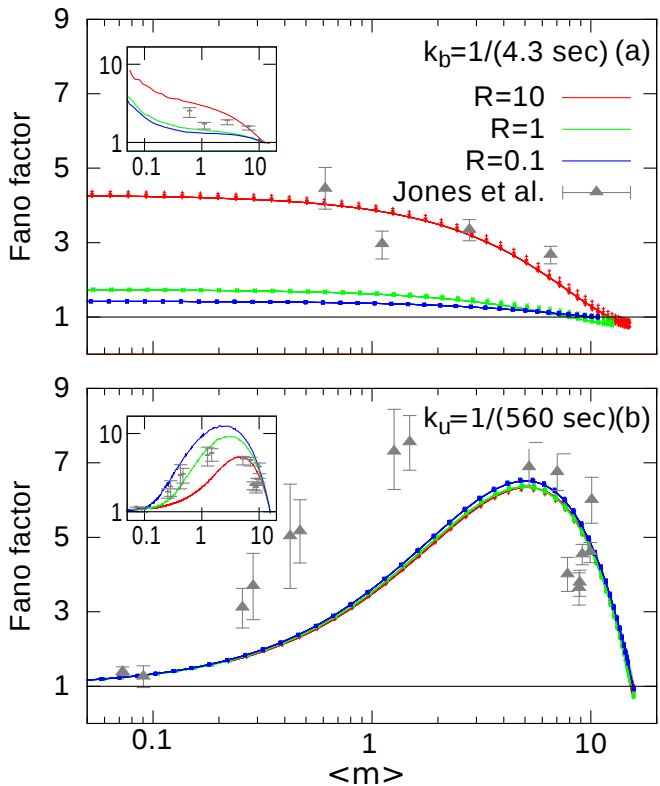}
\caption{
Fano factor as a function of mRNA number $\langle m\rangle$
with $r/\gamma=15.7$, $\gamma=ln(2)/(117sec)$ \cite{petersen}
and $R=10$, $1$, and $0.1$.
Solid lines are for the two-step model (\ref{fano2step}),
while symbols are obtained by three-step model (\ref{fano3step})  with 
combinations of $r_1$, $r_{-1}$, $r_2$, and $r_3$ 
corresponding to the used $R$ and $r$.  
Symbols with error-bars are the corresponding experimental data from
Fig.~3 in Jones {\it et al.} \cite{Jones}.
Insets show the Fano factor vs. $\langle m \rangle$ where the repressor number fluctuation is taken into account by simulating the stochastic production of repressor mRNA and protein, which makes $k_b$ a stochastic variable.
The detail of the simulation is given in Appendix~\ref{Repsimulation}.
(a) Assuming 9.8 tetramers per cell we set 
$k_b=1/(4.3 sec)$.  $k_u$ is varied to change 
$\langle m\rangle$. 
Inset: Effect of cell-to-cell variation of $k_b$ due to stochastic repressor production, with 9.8 tetramers per cell on average.
(b) $k_u=1/(560sec)$ from \cite{hammar} and $k_b$ is varied to change $\langle m\rangle$.
Inset: Effect of cell-to-cell variation of $k_b$, where the average repressor number is controlled through the transcription rate of the repressor mRNA. 
}\label{figure1}
\end{figure}

The Fano factor for the effective two-step initiation model with repression, eq. (\ref{fano2step}), 
becomes
\begin{equation}
\nu\equiv \frac{\sigma^2}{\langle m\rangle}
\approx 1+\frac{(r_{on}/k_b)-R\cdot K\cdot (1+K)}{[1+K(1+R)]^2}. 
\label{nu2stepaprox}
\end{equation}
when mRNA degradation rate  (typically $\sim$ 1/(3 min) \cite{pedersen1978functional})  is much smaller than the repressor binding rate as well as 
the RNAP elongation rate ($\gamma \ll k_b$ and $\gamma \ll r_f$). 

The importance of  the on-rate $r_{on}$ for the cell to cell variability becomes 
evident when we consider the substantially repressed genes,
or  genes 
where the concentration dependent on-rate of the repressor
$k_b$ is much higher than the off rate $k_u$, i.e., $K\to 0$.
In this case, we have  
\begin{equation}
\nu \approx 1 + \frac{r_{on}}{k_b}.
\label{one}
\end{equation}
The increase of $\nu$ with $r_{on}/k_b$ reflects the
number of transcription initiations between each 
repressor binding event  \cite{nakanishi}. 
The difference of eq.~(\ref{one}) from the simple promoter model
\cite{so,Jones} is that noise can be large for a weak promoter in case
its low basal activity is caused by limiting later steps 
in the transcription initiation (e.g. the lac promoter).

Experimentally, the noise is typically measured as a function of
the average mRNA number $\langle m\rangle$ 
\cite{so,sanchez,Jones}.  The average mRNA number can be controlled 
by either changing the repressor binding strength to the promoter 
(typically by altering the binding site sequence on the DNA),
or by changing the concentration of the repressor. The former corresponds 
to changing $k_u$ at a fixed $k_b$, while the latter is the other way around. 

For a fixed repressor concentration (constant $k_b$), 
from (\ref{repressm}) we can express $k_u$ as a function of $\langle m \rangle $.
Replacing $k_u$ in the Fano factor, one get 
\begin{eqnarray}
\nu & \approx &  1+ \frac{r_{on}}{k_b^*} \cdot
\left( 1 - \frac{\langle m \rangle}{m_{max}} \right)^2
\label{two}
\end{eqnarray}
with $m_{max}\equiv r/\gamma$, 
where the approximation ignores a reduction term in Fano factor, which is small when 
$\langle m\rangle < m_{max}$, see eq.~(\ref{nu2stepaprox}).

The prefactor is governed by the $\gamma$-corrected association rate $k_b^*\approx k_b+\gamma (R+1) 
(1-\frac{\langle m\rangle }{m_{max}}) \sim k_b$.
This means that $\nu$ decreases monotonically with $\langle m\rangle$
when the change is caused by increased $k_u$ (by operator mutations). 

As an illustration we now consider the substantial Fano factors 
that was measured on the Lac system by \cite{Jones}. 
A one-step model ($R\ll 1$) would require 
$k_b$ values that are much smaller than the overall initiation rate $r$ ($\sim 1/(11\; sec)$  for the measured Lac system \cite{petersen,Jones}) 
to have a Fano factor substantially larger than 1, because $r_{on}\approx r$ when $R\ll 1$. 
Indeed \cite{Jones} uses the binding rate for one Lac tetramer to be one per 6.3 minutes to fit the measured $\nu$ with a one-step model. 
However, this rate may be too slow given that the association 
rate of one Lac-dimer is estimated to be about 1/3.5 min \cite{hammar2012}, 
and is found to bind 5 fold slower than a Lac-tetramer \cite{hsieh,oehler},
suggesting an association rate per tetramer of 1/42 sec in 
an {\sl E. coli} cell.

The multi-step models can give high Fano factors 
at much higher values of $k_b$.
Spassky {\it et al.} \cite{spassky} measured that open complex formation
takes $1/r_2\sim 10$ sec for the lacUV5 promoter in vitro,
which combined with $r \sim 11$sec  
suggest that this later step is rate limiting and that
$R\gg 1$. 
Our analysis assuming $R=10$ 
and on-rate of a single Lac-tetramer of 1/42 sec gives the Fano factors of $\sim 4$ with $\sim 10$ tetramers per cell, consistent with the experimental data by Jones {\it et al.} \cite{Jones} (Fig.\ref{figure1}a). 

Consider now a given operator (constant $k_u$) 
and change $\langle m \rangle$ by
regulating the repressor concentration ($k_b$). 
The Fano factor in this case is 
\begin{eqnarray}
\nu & \approx & 1+  
\frac{\gamma}{k_u^*} \cdot \langle m \rangle \cdot 
\left( 1-  \frac{ \langle m \rangle }{m_{max}}  \right),
\label{three}
\end{eqnarray}
with the $\gamma$-corrected dissociation rate 
$k_u^*\approx$ $k_u + \gamma \cdot \langle m\rangle / m_{max}$ 
$\approx k_u$.
Eq.~(\ref{three}) is non-monotonic, with largest $\nu$ 
at half maximum expression $\langle m \rangle \sim m_{max}/2$ (Fig.~\ref{figure1}b).
The functional dependence of $\nu$ with $\langle m \rangle $
in Eq.~(\ref{three}) does not depend on $R$,
but the interpretation of the underlying dynamics does.
Noticeably, to obtain a given repression level $m/m_{max}$ 
for a promoter with $R\gg 1$ the repressor needs a factor $(1+R)$ stronger binding
than naively expected.
This reflects that the repressor has to act in the reduced time
where the promoter is not occupied by RNA polymerase \cite{nakanishi}, see Fig. 1. A corollary of this interplay is that 
estimates of repressor binding energies
from promoter activities also rely on the non-equilibrium 
aspects of the RNAP-promoter dynamics.

{%\bf
  Finally one may notice that the fit in Fig.~3b underestimates
the measured noise level. Part of this is attributed to the fact that we use a 
rate $k_b$ that is the same for all cells at a given repression level.
The Lac-tetramer in fact comes at small numbers \cite{semsey2013effect}, 
even at the highly repressed state (to the left of Fig.~3b), and the cell-to-cell
fluctuations can be substantial. This will add to the Fano-factor
and in particular increase the variation of mRNA for intermediate repression levels, and make the maximum Fano-factor to be reached substantially
below the $\langle m \rangle \sim m_{max}/2$ value predicted by 
eq. \ref{three}.

We simulated this effect by considering stochastic production and degradation of the repressor (Fig.~\ref{figure1}b inset). The system is more sensitive to this fluctuation for smaller $R$, because the repressor is more effective (see eq.~\ref{repressm}). In the simulation, we set the parameter so that one 
repressor is produced per one repressor mRNA. 
The effect will be naturally larger if more than four LacI monomers are produced per mRNA. Also the effect will be stronger 
when the chromosome copy number is larger than one, because fluctuations
of repressor act simultaneously on each gene copy. 

For Fig.~\ref{figure1}a, the fluctuation of repressor number also increase the Fano factor
(Fig.~\ref{figure1}a inset), but a larger $R$ shows a larger Fano factor
for the same $\langle m \rangle$. This is because to achieve the same
$\langle m \rangle$, $k_u$ needs to be smaller for larger $R$, which makes 
the dynamics more bursty and sensitive to the repressor number fluctuation. 
}

\section{Discussion}
The above analyses only apply to
repressors that act by simple occlusion, and do not affect the
post binding steps of transcription initiation.
In case a transcriptional repressor acts by stalling the 
isomerization step \cite{roy, roy2}, it does not occlude the RNAP binding site and
the noise should scale with $r$ as suggested by the $R\ll 1$ limit \cite{so}. 
In case the transcriptional regulator is an activator, it may 
act through modification of $r_1$, $r_{-1}$, $r_2$ or $r_3$
\cite{roy} but will not occlude the promoter, and we therefore expect
the burstiness to be reproduced by considering an overall
initiation rate modulation as implied in the formalism of \cite{so}. 

This short paper aimed to clarify
the interplay between time-scales of transcription initiation, and 
time-scales of transcriptional repressors 
in prokaryotes. As an added benefit, the 
formalism propose to use measurements of 
the Fano factor as a tool 
to determine the ratio of two competing 
rates (eq.~(\ref{one})). By exposing 
for example a promoter with large $r_{on}$ to different repressors, 
one may compare repressor dynamics. Conversely, 
by exposing different promoters to the same repressor/operator combination,
one may quantify their relative on-rates for RNAP.
{%\bf
  To be truly useful, such an experimental design should 
preferentially use a repressor which exhibits minimal noise,
as one thereby reduce the extrinsic noise. }

Finally, although Fano factors in principle are robust to having multiple
copies of a given promoter in the E.coli cell, then one should be aware that 
failure in detecting all mRNA will make the experimentally
measured Fano factor systematically smaller 
than the real one,
\begin{equation}
\nu_{measured} = 1+ p \cdot (\nu -1 )
\end{equation}
where $p$ is the probability for observing a mRNA
in the cell (detail in appendix \ref{expmeasure}). 
For instance, the procedures based on counting individual 
spots tend to underestimate the number of mRNA molecules  \cite{skinner2013measuring}; the highest value of mRNA  per cell reported in ref.~\cite{taniguchi}, 
which uses the counting method, is less than 10, while 
ref.~\cite{so} that uses the total intensity to estimate the mRNA number reports $\sim$ 50 mRNAs per cell.  
Thus, if $p$ is say 0.2, then a real burst size
of $\nu \sim 9$ would only be detected as 
$\nu_{measured} \sim 2.6$. Therefore a 
measured Fano factor should be corrected 
by the estimated likelihood for identifying
individual mRNAs in the cell. 

Using $\nu$ as a experimental tool to learn about
promoter dynamics would further be facilitated by reporter mRNAs
with relatively large lifetimes (small $\gamma$).
Central in such an analysis is to realize that
transcriptional noise is primarily sensitive to the first 
steps of the transcription initiation process (Fig.~\ref{figure1}),
and thereby cell to cell variations becomes sensitive to
the limiting process of individual promoters.

\appendix
\section{Derivation of the Fano factor}\label{fanoderive}
We summarize the derivation of the Fano factor for the model described in Fig.~\ref{modelfig}.
In the three-step transcription initiation model, the promoter can be in one of the following four states: free ($f$), RNAP forming a closed complex ($c$), RNAP forming an open complex ($o$), 
and bound by the transcriptional repressor ($T$). In this model repressor binding does not influence open complex formation or the rate of elongation
initiation. 
We denote the probability for the promoter to be in the state $\alpha$
and having $m$ mRNAs at time $t$ to be $P^\alpha_m(t)$, where
$\alpha$ can be $f$, $c$, $o$, or $T$. Assuming that a mRNA is produced
at the moment the RNAP elongates (this ignores the deterministic clearance time), we have the following master equations:
\begin{eqnarray}
  \dot{ P}_m^f(t)&=&r_3 P^o_{m-1}(t)+k_u P^T_{m}(t)+r_{-1} P^c_m(t)-(r_1+k_b)P^f_m(t)\nonumber\\&&
  +\gamma\left[(m+1)P^f_{m+1}(t)-mP^f_{m}(t)\right], \nonumber\\
    \dot{P}_m^c(t)&=&r_{1} P^f_{m}(t)-(r_{-1}+r_2) P^c_{m}(t) \nonumber\\
  &&  +\gamma\left[(m+1)P^c_{m+1}(t)-mP^c_{m}(t)\right],\nonumber\\
    \dot{P}_m^o(t)&=&r_2 P^c_{m}(t)-r_3 P^o_{m}(t)\nonumber\\
   && +\gamma\left[(m+1)P^o_{m+1}(t)-mP^o_{m}(t)\right],\nonumber\\
 \dot{P}_m^T(t)&=&k_b P^f_{m}(t)-k_u P^T_{m}(t)\nonumber\\
&&  +\gamma\left[(m+1)P^T_{m+1}(t)-mP^T_{m}(t)\right].\nonumber
\end{eqnarray}
The probability to have $m$ mRNAs in the system at time $t$ irrespective of the promoter/operator state is given by $P_m(t)\equiv P_m^f(t)+P_m^c(t)+P_m^o(t)+P_m^T(t)$. 
The Fano factor $\nu = (\langle m^2\rangle -\langle m\rangle^2)/\langle m \rangle$ was obtained by calculating 
$\langle m \rangle=\sum_{m=0}^\infty mP_m$ and $\langle m^2 \rangle=\sum_{m=0}^\infty m^2P_m$ in the steady state using the generating function method \cite{vanKampen}. 
The resulting Fano factor for the three-step model is given by
\begin{widetext}
\begin{equation}\nu_{3-step}=1+
%\frac{\frac{r_{on}}{k_b}-R\cdot K\cdot (1+K^*)
 %-\frac{r_{on}^2}{r_2 r_3}KK^*-\frac{r_{on}^2}{r_1 r_3}[1+\frac{\gamma}{r_2}]K(1+K^*)
 %}{[1+K^*(1+R)+\frac{\gamma}{r_3}(1+K^*)(1+\frac{r_{on}\gamma }{r_1 r_2})
 %+K^*\frac{r_{on}\gamma}{r_3r_2}
 % ]\cdot [1+K\cdot (1+R)]}.
\frac{\frac{r_{on}}{k_b}-K(1+K^*)\left(
\frac{r_{on}}{r_3}+\frac{r_{on}^2}{r_1r_2}+\frac{\gamma r_{on}^2}{r_1r_2r_3}\right)-\frac{r_{on}^2}{r_2 r_3}KK^*}
{\left[K^*\left(R+\frac{\gamma r_{on}}{r_2 r_3}\right)+\left(1+\frac{\gamma}{r_3}\right)\left(1+K^*\right)\left(1+\frac{\gamma r_{on}}{r_1r_2}\right)\right]\cdot \left[1+K\cdot (1+R)\right]}.
 \label{fano3step}
 \end{equation}
\end{widetext}
with $K^*\equiv K+\gamma/k_b$ and the on-rate $r_{on}=r_1 \cdot r_2/(r_{-1}+r_2)$ that is modulated from the rate $r_1$ because 
the RNAP may unbind from the promoter.

The Fano factor for the effective two-step initiation model (RNAP binding and elongation initiation) can also be obtained similarly,
or by taking $r_2\to \infty$ limit of eq.~(\ref{fano3step}) noting that $r_{on}\to r_1$ and $r_f\to r_3$ in this limit. 
%In this case the closed complex formation step is ignored and the master equations become
%\begin{eqnarray}
%  \dot{P}_m^f(t)&=&r_f P^o_{m-1}(t)+k_u P^T_{m}(t)-(r_{on}+k_b)P^f_m(t)\nonumber\\&&
%  +\gamma\left[(m+1)P^f_{m+1}(t)-mP^f_{m}(t)\right], \nonumber\\
%    \dot{P}_m^o(t)&=&r_{on} P^f_{m}(t)-r_f P^o_{m}(t)\nonumber\\
%   && +\gamma\left[(m+1)P^o_{m+1}(t)-mP^o_{m}(t)\right],\nonumber\\
% \dot{P}_m^T(t)&=&k_b P^f_{m}(t)-k_u P^T_{m}(t)\nonumber\\
%&&  +\gamma\left[(m+1)P^T_{m+1}(t)-mP^T_{m}(t)\right].\nonumber
%\end{eqnarray}
The full expression of the Fano factor for the effective two-step model is given by
\begin{widetext}
 \begin{equation}
   \nu_{2-step}=1+\frac{(r_{on}/k_b)-R\cdot K\cdot (1+K^*)}{[1+K^*(1+R)+(\gamma/r_f)(1+K^*)]\cdot [1+K\cdot (1+R)]}.
    \label{fano2step}
   \end{equation}
   \end{widetext}

\section{Stochastic fluctuation of the repressor number} \label{Repsimulation}
We assume that the repressor mRNA is transcribed at a constant rate $\alpha$ and
degraded at a rate $\Gamma_m$ per mRNA. Each mRNA is translated
at a rate $\beta$ to produce a repressor and the repressor is degraded at a rate $\Gamma_p$ per repressor, as parametrized in\cite{friedman2006linking}. For simplicity, we assume that a produced repressor
corresponds to a LacI tetramer.
We employ the two-step model in Fig.~\ref{modelfig} for the promoter dynamics,
with making the repressor binding rate dependent on the number of repressor molecules $N_r$ as $k_b=k_0\cdot N_r$, where $k_0=1/$(42 sec) is the single repressor binding rate. For all the simulations, we assumed $\Gamma_m=\ln(2)/(120sec)$,
$\Gamma_p=\ln(2)/(40 min)$, and $\beta=\Gamma_m$, i.e., one repressor tetramer is produced per mRNA on average. For Fig.~\ref{figure1}(a) inset, $\alpha$ is set to be
$9.8\cdot \Gamma_m\cdot \Gamma_p/\beta$ to have 9.8 repressors (tetramers) on average,
while Fig.~\ref{figure1}(b) inset, $\alpha$ is changed to control $\langle m \rangle$. The reactions were simulated by Gillespie method \cite{gillespie1977exact}, and averages were calculated from the data. 

\section{Effect of limited detection on the measured Fano factor}\label{expmeasure}
Suppose when we make the observation, each mRNA can be observed with a constant
probability $p$. When the probability to have $m$ mRNA is $P_m$,
then the probability $Q(n)$ to observe $n$ mRNAs is
\begin{equation}
Q(n)=\sum_{m=0}^\infty  \frac{m!}{n!(m-n)!} p^n (1-p)^{m-n}P_m \Theta (m-n),
\end{equation}
where $\Theta(x)$ is the Heaviside step function. 
This gives
\[
  \langle n \rangle =\sum_{n=0}^\infty n Q(n)=
 p\langle m \rangle, 
\]
and 
\[
\langle n^2\rangle = p^2\langle m(m-1)\rangle+p\langle m\rangle.
\]
This results in the measured Fano factor to be 
\begin{eqnarray}
\nu_{measured}&=&\frac{\langle n^2\rangle-\langle n\rangle^2}{\langle n \rangle}
= p\frac{\langle m^2\rangle-\langle m\rangle^2}{\langle m \rangle}+(1-p)\nonumber \\
&=&1+p\cdot (\nu-1),
\end{eqnarray}
where $\nu$ is the actual value of the Fano factor. 

\begin{acknowledgments}
We tank for support from the Danish National Research Foundation
through the Center for Models of Life.
\end{acknowledgments}

\bibliography{fanofactor}

\end{document}